\documentclass[%        	Class options:
aps,%                   	American Physical Society
prd,%                   	Physical Review D
showpacs,%              	Displays PACS after abstract
twocolumn,
superscriptaddress,%    	Authors' addresses linked with superscripts
nofootinbib,%           	Does not treat footnotes as references
floatfix]%              	Fixes float errors
{revtex4-1}%              	REVTEX 4 Package used
\usepackage{graphicx,%  	Default Latex 2eps package for embedding figures
longtable,%             	Useful for long table
slashed,%						R-parity slash
hyperref,
bm,url,
braket,
xcolor,
xspace}
\usepackage[english]{babel}
\usepackage[utf8]{inputenc}
\usepackage[T1]{fontenc}
\usepackage{amsfonts}
\usepackage{amssymb}
\usepackage{amsmath}
\usepackage{dsfont} 
\usepackage{bm}
\usepackage{braket}
\usepackage{slashed}
\usepackage{simplewick}
\usepackage[top=2cm,bottom=2cm,left=2cm,right=2cm]{geometry}
\usepackage{enumitem}
\usepackage{booktabs}
\usepackage{graphicx}
\usepackage{xcolor}
\begin{document}
%======================================================================================
%:{SK_23.09.2017} Variants of the title.
%\title{The interpolating formula for the $0\nu\beta\beta$-decay half-life including
%  arbitrary neutrino mass  within left-right symmetric models}
%\title{On the possibility of distinguishing light from heavy neutrino contribution to $0\nu\beta\beta$-decay within left-right symmetric models}
%\title{If light and heavy neutrino contributions to $0\nu\beta\beta$-decay are distinguishable experimentally? 
%The case of left-right symmetric models}
%\title{If light and heavy neutrino contributions to $0\nu\beta\beta$-decay are experimentally distinguishable? 
%%The case of left-right symmetric models
%}
% \title{Are light and heavy neutrino contributions to $0\nu\beta\beta$-decay experimentally distinguishable 
%The case of left-right symmetric models

\title{The interpolating formula for the $0\nu\beta\beta$-decay half-life in the case of light and heavy 
neutrino mass mechanisms
}

%:{SK_23.09.2017} END
%--------------------------------------------------------------------------------------
%
\author{A.~Babi\v{c}}
\affiliation{Dept.~of Dosimetry and Application of Ionizing Radiation, Czech Technical U., 115~19~Prague, Czech Rep.}
\affiliation{Institute of Experimental and Applied Physics, Czech Technical University, 128~00~Prague, Czech Republic}
\affiliation{Bogoliubov Laboratory of Theoretical Physics, Joint Institute for Nuclear Research, 141980~Dubna, Russia}
\author{S. Kovalenko}
\affiliation{Universidad T\'{e}cnica Federico Santa Mar\'ia,
Centro-Cientifico-Tecnol\'{o}gico de Valparaiso,
Casilla 110-V, Valparaiso, Chile}
\author{M.I. Krivoruchenko}
\affiliation{Bogoliubov Laboratory of Theoretical Physics, Joint Institute for Nuclear Research, 141980~Dubna, Russia}
\affiliation{Institute for Theoretical and Experimental Physics, B. Cheremushkinskaya 25, 117218 Moscow, Russia}
%
%%
%\author{        S.T.~ Petcov\footnote{Also at: Institute of Nuclear Research and
%Nuclear Energy, Bulgarian Academy of Sciences, 1784 Sofia, Bulgaria}}
%\affiliation{SISSA, Via Bonomea 265, 34136 Trieste, Italy }
%\affiliation{INFN, Sezione di Trieste, 34126 Trieste, Italy}
%\affiliation{Kavli IPMU (WPI), The University of Tokyo, Kashiwa,
%Japan}
%
\author{        F.~\v{S}imkovic}
\affiliation{	Department of Nuclear Physics and Biophysics,
				Comenius University,
				Mlynsk\'a dolina F1, SK--842 15 Bratislava, Slovakia}
\affiliation{	Bogoliubov Laboratory of Theoretical Physics, JINR, %\\
				141980 Dubna, Moscow Region, Russia}
\affiliation{  Czech Technical University in Prague,
   				CZ-12800 Prague, Czech Republic}
\begin{abstract}%.......................................................................
  \vspace*{0cm}
We revisit the ``interpolating formula'' proposed in our previous publication. It allows one to calculate the $0\nu\beta\beta$-decay half-life 
%including
for arbitrary neutrino mass without involvement of the complicated 
results for nuclear matrix elements (NME) obtained within specific nuclear structure models.  
%proposed in our previous publication
%and 
%
%within left-right symmetric models is presented. 
The formula derives from the finding that the value of a properly normalized ratio of the NMEs for the light and heavy neutrino mass mechanisms
%shows 
weakly depends on isotope.  
%Importantly, this holds for all the known nuclear structure approaches. 
%  , which can be
%  interpreted as squared averaged momentum of exchange neutrino, 
%depends very weakly on a given isotope.  
From this fact it follows, in particular,  that the light and heavy neutrino mass mechanisms can hardly be distinguished in a model independent way searching for $0\nu\beta\beta$-decay 
%experiments 
  %observing this rare process on 
of different nuclei. 
Here we show that this formula holds for all the known nuclear structure approaches. 
We give a mathematical justification of our results examining analytical properties of the NMEs.
We also consider several simplified benchmark scenarios within left-right symmetric models 
and 
analyze the conditions for the dominance of the light or heavy neutrino 
mass mechanisms in $0\nu\beta\beta$-decay.
%
%study a lepton number violating parameter taking into account light and heavy neutrino
%contributions to $0\nu\beta\beta$-decay.
%
%% masses is introduced and discussed for simplified neutrino mixing scenarios. 
%
%The general lepton number violating parameter including light and heavy neutrino masses is introduced and discussed for simplified neutrino mixing scenarios. 
\end{abstract}%.........................................................................
%\medskip
%\pacs{%PACS Numbers:
%23.40.-s, 21.60.Jz, 02.50.-r} 
\maketitle
%%%%%%%%%%%%%%%%%%%%%%%%%%%%%%%%%%%%%%%%%%%%%%%%%%%%%%%%%%%%%%%%%%%%%%%%%%%%
%%%%%%%%%%%%%%%%%%%%%%%%%%%%%%%%%%%%%%%%%%%%%%%%%%%%%%%%%%%%%%%%%%%%%%%%%%%%
\section{\label{sec:introduction}Introduction}
%%%%%%%%%%%%%%%%%%%%%%%%%%%%%%%%%%%%%%%%%%%%%%%%%
Neutrinoless double beta  ($0\nu\beta\beta$) decay is a Lepton Number Violating (LNV) process changing lepton number in two units $\Delta L = 2$. It is forbidden in the Standard Model (SM), where lepton number is conserving. Basically there are two sources of LNV:  Majorana neutrino mass and LNV vertices.  The latter may emerge from numerous high-scale models giving rise to the corresponding mechanisms of $0\nu\beta\beta$-decay. 
Once this process is observed, the question of  distinguishing between the dominant mechanisms will arise.  Certainly, this task is highly non-trivial. One may hope that measurements of  $0\nu\beta\beta$ half-life with different isotopes would facilitate its solution due to variability of  Nuclear Matrix Elements (NME) of particular  mechanisms from one isotope to the other. 
In the present paper we show that at least the light and heavy Majorana neutrino mass mechanisms are indistinguishable 
in this way without additional hypothesis. This fact becomes especially comprehensible in terms of the so called ``interpolating formula'' (IntF)\cite{sterile14} merging the light and heavy neutrino mass ranges in the NMEs and allowing a transparent physical interpretation of the above fact. 
The IntF is a simple analytical formula representing with an accuracy of 30\% or better the NME as a function of the Majorana neutrino mass. This accuracy is sufficient for practical purposes taking into account the limited accuracy of the available nuclear structure approaches to the NME calculations. 
In what follows we will show that that the IntF is valid for all these nuclear structure approaches with the above-indicated accuracy and elucidate some of its other useful properties.
On the particle physics side we adopt a generic scenario with Majorana neutrinos of arbitrary value masses and consider their contribution to $0\nu\beta\beta$-decay via mass mechanism mediated by both left- and right-handed weak currents.  
Then for the sake of concreteness we consider 
the neutrino mass mechanism within the left-right symmetric models (LRSM) \cite{Pati,Mohap} and extent our analysis towards some more particular scenarios.

%%%%%%%%%%%%%%%%%%%%%%%%%%%
%%%%%%%%%%%%%%%%%%%%%%%%%%%%%%%%%%%%%%%%%%%%%%%%%%%%%%%%%%%%%%%%%%%%%%%%%%%%
%%%%%%%%%%%%%%%%%%%%%%%%%%%%%%%%%%%%%%%%%%%%%%%%%%%%%%%%%%%%%%%%%%%%%%%%%%%%
%%%%%%%%%%%%%%%%%%%%%%%%%%%%%%%%%%%%%%%%%%%%%%%%%%%%%%%%%%%%%%%%%%%%%%%%%%%%
%\section{\label{sec:} $0\nu\beta\beta$-decay within the LRSM}
%%%%%%%%%%%%%%%%%%%%%%%%%%%%%%%%%%%%%%%%%%%%%%%%%%%%%%%%%%%%%%%%%%%%%%%%%%%%

\section{Neutrino mass mechanism of $0\nu\beta\beta$-decay}
\label{sec:} 
%%%%%%%%%%%%%%%%%%%%%%%%%%%%%%%%%%%%%%%%%%%%%%%%%%%%%%%%%%%%%%%%%%%%%%%%%%%%
We start with a generic Majorana neutrino mass mechanism of $0\nu\beta\beta$-decay induced by the low-energy effective Lagrangian
%
%The LRS models  extend the Standard Model (SM) gauge symmetry
%to the group \mbox{$SU(2)_L\otimes SU(2)_R \otimes U(1)_{B-L}$} with an additional neutral
%$Z'$ and two charged vector bosons $W^\pm_R$. This symmetry group is broken spontaneously down to the SM group 
%at a sufficiently high energy scale $\Lambda_{LR}$ providing to the extra gauge bosons $Z', W^\pm_R$ large masses 
%$\sim \Lambda_{LR}$.
%After integrating out these heavy particles one ends up at low energies  with the well known current-current effective Lagrangian involving both left- and right-handed weak currents. The part,  
%which can trigger the $0\nu\beta\beta$-decay is \cite{Doies}
%
\begin{eqnarray} \label{hamilweakrh}
\mathcal{L}^\beta &=& \frac{G_{\beta}}{\sqrt{2}} ~\left[
j_L^{~\rho}J^{\dagger}_{L\rho }~+~\lambda j_R^{~\rho}J^{\dagger}_{R\rho } +  h.c.\right]
\end{eqnarray}
with the left$/$right-handed hadronic $J_{L/R}$ and leptonic  $j_{L/R}$ currents. As usual, $G_{\beta}=G_F\cos{\theta_C}$, where $G_F$ and $\theta_C$ are Fermi constant and Cabbibo angle, respectively.  The dimensionless parameter $\lambda$ depends on the underlying high-scale model. In the particular case of  the Left Right Symmetric (LRS) models, based on the \mbox{$SU(2)_L\otimes SU(2)_R \otimes U(1)_{B-L}$} gauge group \cite{Pati,Mohap}, the Lagrangian (\ref{hamilweakrh}) appears at low energies after integrating out 
$W^{\pm}_{L,R}$ massive gauge bosons. In this model 
%%We will assume the dominance of  the neutrino mass mechanisms in $0\nu\beta\beta$-decay originating from the mass term in the neutrino propagator connecting two vertices both with $W_L$ (standard mechanism) as well as both with $W_R$ ($\lambda^2$ mechanism). 
%This is motivated by the recent analysis \cite{now16} of the mixed $W_L-W_R$-contribution 
%($\lambda$ mechanism) via the momentum dependent part of the neutrino propagator, which indicates its subdominant role in $0\nu\beta\beta$-decay. 
%For sake of simplicity the possible $W_L - W_R$-mixing is not taken into account and
%the coupling constants $\lambda$ is chosen to be real 
%
\begin{eqnarray} \label{relacieLRSM}
\lambda =   (M_{W_L}/M_{W_R})^2,
\end{eqnarray}
where $M_{W_{L}}$ and $M_{W_{R}}$ ($M_L<M_R$) are masses of $W_L$ and $W_R$ gauge bosons, respectively.
The current constrain on the mass of $W_R$ is $M_{W_R} \ge$ 2.9 TeV \cite{Dev} sets the limit 
\begin{eqnarray}\label{eq:lambda-limit}
&&\lambda \le 7.7 \times 10^{-4}.
\end{eqnarray}
The upper limit $\lambda = 7.7 \times 10^{-4}$ we use everywhere in the present paper as a reference value for this parameter. Since we focus on the mass mechanism we discarded in Eq.~(\ref{hamilweakrh})  the 
$j_{L,R} J_{R,L}$ terms irrelevant in this case
%leading to a subdominant contribution to $0\nu\beta\beta$-decay \cite{now16} 
(for a review see, for instance \cite{Ve12}).  
%non-vanishing in the limit of zero neutrino mass. This the so called $\lambda$-mechanism is mediated by 
%the momentum dependent part of the neutrino propagator connecting 
%.  Note also that in the LRS models 
%We will assume the dominance of  the neutrino mass mechanisms in $0\nu\beta\beta$-decay originating from the mass term in the neutrino propagator connecting two vertices both with $W_L$ (standard mechanism) as well as both with $W_R$ ($\lambda^2$ mechanism). 
%This is motivated by the recent analysis \cite{now16} of the mixed $W_L-W_R$-contribution 
%($\lambda$ mechanism) via the momentum dependent part of the neutrino propagator, which indicates its subdominant role in $0\nu\beta\beta$-decay. 
%
In Eq.~(\ref{hamilweakrh}) the explicit form of the left- and right-handed hadronic currents $J^{\dagger}_{L,R}$
in nuclei 
%$J^{\dagger}_{R\rho }$ 
can be found, e.g., in Ref. \cite{Stef15}. The leptonic currents are given by 
\begin{eqnarray}
j_L^{~\rho}=\bar{e}_{L}\gamma^{\rho}\nu'_{eL}, \qquad j_R^{~\rho}=\bar{e}_{R}\gamma^{\rho}\nu'_{eR}.\nonumber 
\end{eqnarray}
The $\nu'_{eL}$ and  $\nu'_{eR}$ are the weak eigenstate electron neutrinos, which are expressed as superpositions of the light and heavy Majorana  mass eigenstate neutrinos  $\nu_j$ and $N_k$
as
%, respectively. The electron neutrino eigenstates can be expressed as
\begin{eqnarray}\label{lambdaeta}
  &&\nu'_{eL}=\sum_{j=1}^3 U_{ej}\nu_{j}  + \sum_{k=1}^{n} S_{ek} N^{C}_{k}, \nonumber \\
  &&\nu'_{eR}=\sum_{j=1}^3 T_{ej}^* \nu^C_{j} + \sum_{k=1}^{n} V_{ek}^*N_{k},
\end{eqnarray}
where the unitary matrix
%The $3\times 3$ block matrices in  flavor space $U, S, T, V$ form a $6\times6$ unitary neutrino mixing matrix \cite{Xing}
%
\begin{eqnarray}
\mathcal{U}&=& \left(
\begin{array}{ll}
U & S\\
 T &V \\
 \end{array}
\right).
\label{maticazmies}
\end{eqnarray}
is the generalizations of the
Pontecorvo-Maki-Nakagawa-Sakata (PMNS) matrix, which 
diagonalizes the general $(3+n)\times (3+n)$ neutrino mass matrix
\begin{eqnarray}
 \mathcal{M} &=& \left(
\begin{array}{ll}
M_L & M_D \\
 M_D^T & M_R \\
 \end{array}
\right)
\label{maticanut}
 \end{eqnarray}
in the basis $(\nu^{\prime}_{e L}, \nu^{\prime}_{\mu L}, \nu^{\prime}_{\tau L}, 
N_{1 R}^{\prime C}, ..., N_{n R}^{\prime C})$. Here $M_{L,R}$ and $M_D$ are Majorana and Dirac mass terms, respectively. 
After diagonalization one should end  up with 3 light $\nu_i$ (i=1,2, 3) and n heavy $N_k$ (k=1,...,n) Majorana neutrino mass eigenstates with the masses $m_{i}$ and $M_{k}$, respectively. In the LRS models $n=3$. The smallness of $m_{i}$ can be guaranteed by the seesaw-I condition 
$M_{R}\gg M_{D}$. As is well known this leads to very heavy states $N_{k}$ with masses 
$M_{k} \gg 1$~TeV being beyond the experimental reach. In the scenarios with $n>3$ the inverse seesaw mechanism can be implemented. In this case among $N_{k}$, accompanying the light 
$\nu_{i}$ states, there can appear moderately heavy or even light Majorana states. Actually, their masses can be of arbitrary value. This is the case of our particular interest, for which we designed the above-mentioned interpolating formula.  

Assuming the dominance of the mass mechanism
%We will assume the dominance of  the neutrino mass mechanisms in $0\nu\beta\beta$-decay originating from the mass term in the neutrino propagator connecting two vertices both with $W_L$ (standard mechanism) as well as both with $W_R$ ($\lambda^2$ mechanism). 
%This is motivated by the recent analysis \cite{now16} of the mixed $W_L-W_R$-contribution 
%($\lambda$ mechanism) via the momentum dependent part of the neutrino propagator, which indicates its subdominant role in $0\nu\beta\beta$-decay. 
%
we write down the $0\nu\beta\beta$-decay half-life
%For the $0\nu\beta\beta$-decay half-life we get
%
\begin{eqnarray}
&&[T_{1/2}^{0\nu}]^{-1} = G^{0\nu} g_{\rm A}^{4} m_p^2  \times \\
  &&\left( \left|\sum_{j=1}^3 U^{2}_{ej}~ m_{j} ~ {M}^{\prime\, 0\nu}_{LL}(m_{j})
  ~+~ \sum_{k=1}^n   S^{2}_{ek}~ M_{k}~  {M}^{\prime\, 0\nu}_{LL}(M_{k})
  \right|^{2} \right.\nonumber\\
     \nonumber 
&& + \lambda^2 \times \nonumber\\
&& \left.\left|\sum_{j=1}^3 T_{ej}^2~ m_{j} ~ {M}^{\prime\, 0\nu}_{RR}(m_{j})
~+~ \sum_{k=1}^n V^{2}_{ek}~ M_{k}~  {M}^{\prime\, 0\nu}_{RR}(M_{k})
  \right|^{2}\right).\nonumber
\label{eq:t12}   
\end{eqnarray}
The proton mass is denoted by $m_{\rm p}$ and $g_{\rm A}$ is the unquenched value of axial-vector
coupling constant ($g_{\rm A} = 1.269$). The phase-space factor $G^{0\nu}$ is tabulated
for various $0\nu\beta\beta$-decaying nuclei in Ref. \cite{phaseint}. 
The NMEs ${M}^{\prime 0\nu}$ as functions of neutrino mass $m_{\nu}$ ($m_{\nu} = m_i$ or $M_k$)
are given by \cite{sterile14}
\begin{eqnarray}
\label{eq:MnuN}
&&{M}^{\prime\, 0\nu}_{LL,RR}(m_{\nu})
= \frac{1}{m_{\rm p}m_{\rm e}}~
\frac{R}{2 \pi^2 g^2_A} \sum_{n} \!\! 
 \int \! d^3x \, d^3y \,  d^3p \nonumber\\
&&\times e^{i\mathbf{\rm p}\cdot (\mathbf{x}-\mathbf{y})} \frac{\bra{0^+_F} 
{J}^{\mu\dag}_{L,R} (\mathbf{x})
\ket{n}\bra{n}
{J}^\dag_{L,R~ \mu} (\mathbf{y}) \ket{0^+_I}}{\sqrt{p^2+m_{\nu}^2} 
(\sqrt{p^2+m_{\nu}^2} + E_n-\frac{E_I-E_F}{2})}  \,. 
\nonumber\\
\end{eqnarray}
Here, $R$ and $m_{e}$ are the nuclear radius 
and the mass of electron, respectively.  We use as usual  $R=r_0 A^{1/3}$ with $r_0=1.2$ fm.  
Initial and final nuclear ground states with energies 
$E_{I}$ and $E_{F}$ are denoted by $\ket{0^+_I}$ and $\ket{0^+_F}$, respectively. 
The summation runs over intermediate nuclear states $\ket{n}$ with energies $E_{n}$.
The weak one-body nuclear charged current  ${J}_{L,R}$ \cite{Stef15,sterile14} depends on the
effective value of axial-vector coupling constant $g^{\rm eff}_{\rm A}$ of the nucleon, 
which is renormalized to a smaller, the so called, ``quenched'' value, 
$g^{\rm eff}_A$ \cite{jvrev16}. 
%In the leading order of non-relativistic approximation
%${J}^{\dag}_{L,R \mu}$ in Eq. (\ref{eq:MnuN}) can be replaced with ${J}^{\dag}_{\mu}$
%given in \cite{sterile14}.
%We note that in the expression for the $0\nu\beta\beta$-decay half-life in (\ref{eq:t12})
%the subdominant left-right interference term has been neglected. 
%%%%%%%%%%%%%%%%%%%%%%%%%%%%%%%%%%%%%%%%%%%%%%%%%%%%%%%%%%%%%%%%%%%%%%%%%%%%
%%%%%%%%%%%%%%%%%%%%%%%%%%%%%%%%%%%%%%%%%%%%%%%%%%%%%%%%%%%%%%%%%%%%%%%%%%%%
\section{\label{sec:mechanisms} ``Interpolating'' formula for the $0\nu\beta\beta$-decay half-life}
%%%%%%%%%%%%%%%%%%%%%%%%%%%%%%%%%%%%%%%%%%%%%%%%%%%%%%%%%%%%%%%%%%%%%%%%%%%%
%%%%%%%%%%%%%%%%%%%%%%%%%%%%%%%%%%%%%%%%%%%%%%%%%%%%%%%%%%%%%%%%%%%%%%%%%%%%

\begin{table*}[!t]
  \caption{The values of the parameter $\sqrt{\langle p^2 \rangle}$ of the interpolating formula
    (\ref{eq:interpol-form-1}),  (\ref{eq:p-def-1})
    for a given isotope and their average value
    $\sqrt{\langle p^2  \rangle_a}$ used in Eq. (\protect\ref{eq:LNV-Mass}) 
     with the variance  $\sigma$ (in parentheses) calculated within different nuclear structure
    approaches: interacting shell model (ISM) (Strasbourg-Madrid (StMa)\protect\cite{StMa09}
and Central Michigan University (CMU)\protect\cite{CMU16} groups),
interacting boson model (IBM)\protect\cite{IBM15},
quasiparticle random phase approximation (QRPA) (Tuebingen-Bratislava-Caltech 
(TBC)\protect\cite{TBC13,Fang15} and Jyv\"askyla (Jy)\protect\cite{QJy15} groups), 
projected Hartree-Fock Bogoliubov approach (PHFB)\protect\cite{phfb12H}, and covariant
density functional theory (CDFT) \cite{ring17}.
The Argonne, CD-Bonn and UCOM two-nucleon short-range correlations  are taken into account.
The non-quenched value of weak axial-vector coupling $g_A$ is assumed.
\label{table.1}}
\centering 
\renewcommand{\arraystretch}{1.1}    
\begin{tabular}{lcccccccccccccccccc}\hline\hline
 Method  & $g_A$ &  src & \multicolumn{11}{c}{$\sqrt{\langle p^2 \rangle}$ [MeV]} & & &  $\sqrt{\langle p^2 \rangle_a}$ ($\sigma$) [MeV]  \\ \cline{4-15}
%  \cline{17-18}
  &  &    & ${^{48}}$Ca  & ${^{76}}$Ge  & ${^{82}}$Se  & ${^{96}}$Zr & ${^{100}}$Mo & ${^{110}}$Pd
                           & ${^{116}}$Cd & ${^{124}}$Sn & ${^{128}}$Te & ${^{130}}$Te & ${^{136}}$Xe & ${^{150}}$Nd & & 
 & & \\ \hline \cline{4-9}  \cline{10-15}
  ISM-StMa   & 1.25 &  UCOM   &   178    &    150     &     149    &             &            &
                              &          &    160     &            &    161      &     159    &          & &  160(10)  \\
ISM-CMU    & 1.27 & Argonne   &   178    &    134     &     138    &             &            &          
                              &          &    153     &            &    159      &     170    &          & &  155(17)  \\
           &      & CD-Bonn   &   203    &    165     &     162    &             &            &
                              &          &    177     &            &    184      &     197    &          & &  181(17)  \\
IBM        & 1.27 & Argonne   &   113    &    103     &     103    &    129      &     136    &      135
                              &   130    &    109     &     109    &    109      &     107    &      155 & &  120(17)  \\
QRPA-TBC   & 1.27 & Argonne   &   189    &    163     &     164    &    180      &     174    &      166
                              &   157    &    186     &     178    &    180      &     183    &          & &  175(11)  \\
           &      & CD-Bonn   &   231    &    193     &     194    &    211      &     204    &      194
                              &   182    &    214     &     207    &    209      &     211    &          & &  205(13)  \\
QRPA-Jy    & 1.26 & CD-Bonn   &          &    191     &     192    &    217      &     207    &      187
                              &   177    &    202     &     196    &    201      &     175    &          & &  194(13)  \\
PHFB       & 1.25 & Argonne   &          &            &            &    130      &     127    &      124
                              &          &            &    131     &    132      &            &      121 & &  128(4)  \\
           &      & CD-Bonn   &          &            &            &    150      &     145    &      143
                              &          &            &    150     &    150      &            &     139  & &  146(5)  \\ 
CDFT       & 1.25 & Argonne   &   122    &     129    &    131     &    129      &     131    &      
                              &   133    &     138    &            &    138      &     137    &     138 & &  132(5) 
\\ \hline \hline
\end{tabular}
\end{table*}
For the Majorana neutrino exchange mechanism in the literature there usually considered two limiting cases: 
light $m_{i}\ll p_{\rm F}$ and heavy  $M_{i} \gg p_{\rm F}$
neutrinos, where $p_{\rm F}\sim$ 200 MeV is the Fermi momentum.
For these limiting cases the half-life formula (\ref{eq:t12}) is reduced to:
\begin{eqnarray}
&&[T_{1/2}^{0\nu}]^{-1} = G^{0\nu} g_{\rm A}^{4}\times \nonumber\\  
&&\times \left\{\begin{array}{ll}
  \left|\eta_\nu\right|^{2} 
     \left|{M}^{\prime 0\nu}_{\nu}\right|^{2},  
& \mbox{for}\  m_{i} \ll p_{\rm F},    \\[3mm]
\left|\eta_{\rm N}\right|^{2} 
   \left|{M}^{\prime 0\nu}_{\rm N}\right|^{2}, 
    &  \mbox{for}\  M_{k} \gg p_{\rm F},  
\end{array}\right.
\label{LightHeavy}   
\end{eqnarray}
with 
\begin{eqnarray}\label{meanMass}
  |\eta_\nu|^2 ~ m^2_e &=&
  \left|\sum_{j=1}^3 U^{2}_{ej}~ m_{j} \right|^2
+ \lambda^2 \left|\sum_{j=1}^3 T^{2}_{ej}~ m_{j} \right|^2
\nonumber\\
&\simeq&
  \left|\sum_{i=1}^3 U^{2}_{ej}~ m_{j} \right|^2
\nonumber\\  
|\eta_N|^2 ~\frac{1}{m^2_p} &=& \left(|\eta^L_N|^2 + |\eta^R_N|^2\right) 
~\frac{1}{m^2_p} \nonumber\\ 
&=& 
  \left|\sum_{k=1}^n S^{2}_{ek}~\frac{1}{ M_{k}} \right|^2
  + \lambda^2 \left|\sum_{k=1}^n V^{2}_{ek}~ \frac{1}{M_{k}} \right|^2
\end{eqnarray}
Here the NMEs ${M}^{\prime 0\nu}_{\nu}, {M}^{\prime 0\nu}_{N}$ are derived from the NME 
${M}^{\prime 0\nu}$ in Eq. (\ref{eq:MnuN}) in the following way
\begin{eqnarray}\label{lim-rel-1}
{M}^{\prime 0\nu} (m_{i}\rightarrow 0)
&=&  \frac{1}{m_{\rm p} m_{\rm e}} {M}^{\prime 0\nu}_{\nu}
,\\
\label{lim-rel-2}
{M}^{\prime 0\nu} (M_{\rm i}\rightarrow \infty )
 & = & \frac{1}{M_{\rm i}^{2}} {M}^{\prime 0\nu}_{\rm N}.
\end{eqnarray}

\begin{figure}
\begin{center}
   \includegraphics[height = 5.8cm]{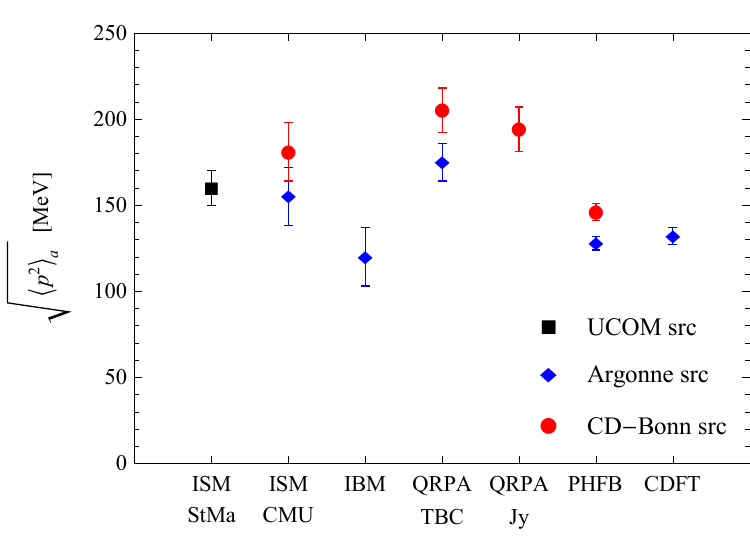}
  \end{center}
\caption{The average value $\sqrt{\langle p^2 \rangle_a}$ over the set of the considered isotopes 
with variance  $\sigma$ calculated within different nuclear structure approaches. The notations are the same as in Table \ref{table.1}. 
\label{fig.paver}}
\end{figure}

In the case of  a neutrino spectrum with mass states $N_{k}$ of  an arbitrary mass value  one has to apply Eq.~(\ref{eq:MnuN}) for the NME calculations, resulting in a complicated function of the neutrino mass. This is a real hassle for use in practice. Fortunately, there is a very good approximate analytical representation for Eq.~(\ref{eq:MnuN}) proposed in 
%
%As it was shown in 
Ref.~\cite{sterile14} (and references therein) and having a remarkably simple form  
%these limiting-case NMEs allow one
%to approximate NME ${M}^{\prime 0\nu}_{\rm LL,RR}(m)$ by the formula
\begin{eqnarray}\label{eq:interpol-form-1}
  {M}^{\prime\, 0\nu}_{LL,RR}(m_{\nu}) \simeq  {M}^{\prime 0\nu}_{\rm N}
  \frac{1} {\langle p^{2}\rangle + m^{2}_{\nu}}
\end{eqnarray} 
This is what we call ``interpolating formula'' (IntF) since it interpolates two limiting cases 
(\ref{lim-rel-1}), (\ref{lim-rel-2}) and is valid to a good accuracy for an arbitrary value of 
$m_{\nu}$. 
Eq.~(\ref{eq:interpol-form-1}) contains the parameter
%with
\begin{equation}
\label{eq:p-def-1} 
\langle p^{2}\rangle = m_{\rm p} m_{\rm e} \frac{M^{\prime 0\nu}_{\rm N}}{M^{\prime 0\nu}_{\nu}}
\end{equation}
with the dimension of (mass)$^{2}$. The form of  Eq.~(\ref{eq:interpol-form-1}) suggests the interpretation of $\langle p^{2}\rangle $ as the mean square momentum of the virtual neutrino propagating between two $\beta$-decaying nucleons. Therefore, it is expected to be of the order of   
$p_{F}^{2} \sim (200\, \rm MeV)^{2}$. 
%nucleon momentum in a nucleus. 
The current values of the matrix elements 
$M^{\prime 0\nu}_{\nu}$ and $M^{\prime 0\nu}_{\rm N}$
calculated within different nuclear structure approaches can be found in Tables
6 and 7 of Ref. \cite{jvrev16}. The value of corresponding parameter $\sqrt{\langle p^{2}\rangle}$ is
given for various isotopes together with its averaged value $\sqrt{\langle p^{2}\rangle_a}$ 
with variance $\sigma$ in Table \ref{table.1}. The unquenched value of axial-vector
coupling constant is assumed: $g^{\rm eff}_A = g_A = 1.25-1.27$. We see that
the value of $\sqrt{\langle p^{2}\rangle}$ depends noticeable
on the chosen nuclear structure method and considered choice of two-nucleon short-range
correlation function. 
The values of $\sqrt{\langle p^{2}\rangle_a}$ are displayed for different nuclear
structure approaches and types of two-nucleon short-range correlations in Fig. \ref{fig.paver}.
The largest value of the parameter $\sqrt{\langle p^{2}\rangle_a} \simeq$ 200 MeV 
is found for the QRPA with isospin restoration and CD-Bonn two-nucleon short range correlations.
Surprisingly, within all the considered 
nuclear structure approaches 
the variance $\sigma$ is very small being of the order of 3-10 \%, i.e. the value of 
$\langle p^{2}\rangle$ is {\it practically the same for all isotopes of experimental interest} and
can be replaced with averaged value $\langle p^{2}\rangle_a$. 
In Appendix we discuss this finding from the view point of the analytical properties of the NME in Eq.~(\ref{eq:MnuN}) as a function in the complex plane of  $m_{\nu}$.
The above conclusion is also supported 
by the statistical treatment of $M^{\prime 0\nu}_{\nu}$ and $M^{\prime 0\nu}_{\rm N}$ NMEs performed in 
Ref.~\cite{lisi15}. 
%Thus we conclude that the light and heavy neutrino exchange mechanisms 
%are indistinguishable from the phenomenological viewpoint. 

Using the parameter $\langle p^{2}\rangle_a$ in the ``interpolating formula'' (\ref{eq:interpol-form-1})
we can write to a good accuracy the $0\nu\beta\beta$-decay half-life 
for the Majorana neutrino exchange mechanism as
% the LRS models can be written as
%By considering the parameter $\langle p^{2}\rangle_a$ the ``interpolating formula''
%for the $0\nu\beta\beta$-decay half-life 
%for Majorana exchange mechanism of the LRS models can be written as
%
\begin{eqnarray}\label{interpol}
  [T_{1/2}^{0\nu}]^{-1} =  \eta_{\nu N}^2 ~C_{\nu N},
\end{eqnarray}
where 
\begin{eqnarray}
\label{coeff-1}
C_{\nu N} \ \ &=& g_{\rm A}^{4} ~\left|M^{\prime 0\nu}_{\nu} \right|^{2} ~G^{0\nu}. 
\label{coeff-2}
\end{eqnarray}
and 
\begin{eqnarray}
\label{eq:LNV-Mass}
&&\eta_{\nu N}^2 = 
\left|\sum_{j}^3 U^{2}_{ej}~ \frac{m_{j}}{m_e} 
  ~+~  \sum_{k}^n S^{2}_{ek}~ \frac{\langle p^2\rangle_a}{\langle p^2\rangle_a + M_k^2}
  ~ \frac{M_{k}}{m_e} 
  \right|^{2} \nonumber \\
 &&~~~~~ + \lambda^2 \left|\sum_{j}^3 T_{ej}^2 ~\frac{m_{j}}{m_e} 
 ~+~ \sum_{k}^n V^{2}_{ek}~ \frac{\langle p^2\rangle_a}{\langle p^2\rangle_a + M_k^2}
 ~ \frac{M_{k}}{m_e}  \right|^{2} \nonumber \\
\end{eqnarray}
for arbitrary mass $M_{k}$. The sum runs over $j=1,2,3$ and $k=1,...,n$. 
The values of parameter $C_{\nu N}$ are given for various isotopes in Table \ref{table.2}.
The ``interpolating formula'' in Eq. (\ref{interpol}) reproduces 
the ``exact'' QRPA result with rather good accuracy except for the transition region where its deviation, as seen from Fig.~\ref{Fig15}, amounts
20\% - 25\%. 
%\cite{sterile14}. 
The parameter $\eta_{\nu N}$ is a general 
LNV parameter for the light and heavy neutrino mass mechanisms,
%in the context of the LRS model, 
which is practically independent of  the isotope under consideration.

\section{Light vs Heavy neutrino mass mechanisms}
\label{sec:LH-mass}

From the conclusion of the previous section and Eq.~(\ref{eq:LNV-Mass}) it follows that contrary to the previous expectations in the literature (see for instance Refs.~\cite{FMPSV11,Meroni13})
the dominance of light or heavy neutrino mechanisms of $0\nu\beta\beta$-decay 
cannot be 
%established 
recognized just by observation of this process with different isotopes. 
An additional theoretical or experimental input 
%or model assumptions concerning 
about neutrino masses and mixing is
needed to shed light on the particular role of each of these mechanisms. 

\begin{table*}[!t]
  \caption{The value of the parameter $C_{\nu N}$ 
%  of the interpolating formula specified 
    in Eq. (\protect\ref{coeff-2}) for the isotopes of experimental interest. The calculated 
light neutrino exchange NME ${M'}^{0\nu}_\nu$ within the
interacting shell model (ISM) (Strasbourg-Madrid (StMa)\protect\cite{StMa09}
and Central Michigan University (CMU)\protect\cite{CMU16} groups),
interacting boson model (IBM)\protect\cite{IBM15},
quasiparticle random phase approximation (QRPA) (Tuebingen-Bratislava-Caltech 
(TBC)\protect\cite{TBC13,Fang15} and Jyv\"askyla (Jy)\protect\cite{QJy15} groups),
projected Hartree-Fock Bogoliubov approach (PHFB)\protect\cite{phfb12H}
and  covariant density functional theory (CDFT) \cite{ring17} are considered.
The Argonne, CD-Bonn and UCOM two-nucleon short-range correlations  are taken into account.
The non-quenched value of the weak axial-vector coupling $g_A$ is assumed.
\label{table.2}}
\centering 
\renewcommand{\arraystretch}{1.1}    
\begin{tabular}{lccccccccccccccccc}\hline\hline
 Method  & $g_A$ &  src & \multicolumn{11}{c}{$C_{\nu N}$ ($10^{-14}$ yrs$^{-1}$) } & & \\ \cline{4-15}
%  \cline{17-18}
  &  &    & ${^{48}}$Ca  & ${^{76}}$Ge  & ${^{82}}$Se  & ${^{96}}$Zr & ${^{100}}$Mo & ${^{110}}$Pd
                           & ${^{116}}$Cd & ${^{124}}$Sn & ${^{128}}$Te & ${^{130}}$Te & ${^{136}}$Xe & ${^{150}}$Nd & & 
 & \\ \hline
  ISM-StMa   & 1.25 &  UCOM   &   4.38   &    4.56    &     17.3   &             &            &
                              &          &    15.1    &            &    24.4     &     17.1   &          & \\
ISM-CMU    & 1.27 & Argonne   &   4.12   &    6.96    &     26.8   &             &            &          
                              &          &    9.38    &            &    11.8     &     10.0   &          & \\
           &      & CD-Bonn   &   4.98   &    7.81    &     30.3   &             &            &
                              &          &    10.8    &            &    13.7     &    11.7    &          & \\
IBM        & 1.27 & Argonne   &   19.7   &    13.4    &     36.7   &    42.7     &    73.5    &      20.5
                              &   41.6   &    23.9    &     2.56   &    50.5     &    35.2    &      27.0 & \\
QRPA-TBC   & 1.27 & Argonne   &   1.88   &    16.3    &     56.7   &    39.5     &    120.    &      41.4
                              &   70.7   &    15.4    &     3.17   &    55.8     &   18.0     &          & \\ 
           &      & CD-Bonn   &   2.24   &    19.0    &     66.4   &    46.8     &   141.     &      48.9
                              &   81.6   &    19.9    &     3.93   &    70.4     &    22.9    &          & \\ 
QRPA-Jy    & 1.26 & CD-Bonn   &          &    16.5    &    35.6    &    51.1     &   61.0     &      51.6
                              &   76.4   &    64.0    &    3.59    &    57.3     &   31.1     &          & \\ 
PHFB       & 1.25 & Argonne   &          &            &            &    40.5     &   132.     &      59.6
                              &          &            &    2.18    &    50.4     &            &      23.7 & \\ 
           &      & CD-Bonn   &          &            &            &    44.6     &   143.     &     64.7
                              &          &            &    2.39    &   55.0      &            &     25.6  & \\
CDFT       & 1.25 & Argonne   &   47.3   &   22.4     &    74.0    &   216.      &  173.      &     
                              &   128.   &   42.3     &            &   88.2      &  68.0      &      113.  & \\ 
\hline \hline
\end{tabular}
\end{table*}

%%%%%%%%%%%%%%%%%%%%%%%%%%%%%%%%%%%%%%%%%%%%%%%%%%%%%%%%%%%%%%%%%%%%%%%%%%%%
%%%%%%%%%%%%%%%%%%%%%%%%%%%%%%%%%%%%%%%%%%%%%%%%%%%%%%%%%%%%%%%%%%%%%%%%%%%%
%\section{Analysis of the $\eta_{\nu N}$ LNV parameter}
%\subsection{Particular model setups}
%\label{sec:analysis}
%%%%%%%%%%%%%%%%%%%%%%%%%%%%%%%%%%%%%%%%%%%%%%%%%%%%%%%%%%%%%%%%%%%%%%%%%%%%
%%%%%%%%%%%%%%%%%%%%%%%%%%%%%%%%%%%%%%%%%%%%%%%%%%%%%%%%%%%%%%%%%%%%%%%%%%%%

\begin{figure}[t]
\centering
\includegraphics[width=0.5\textwidth]{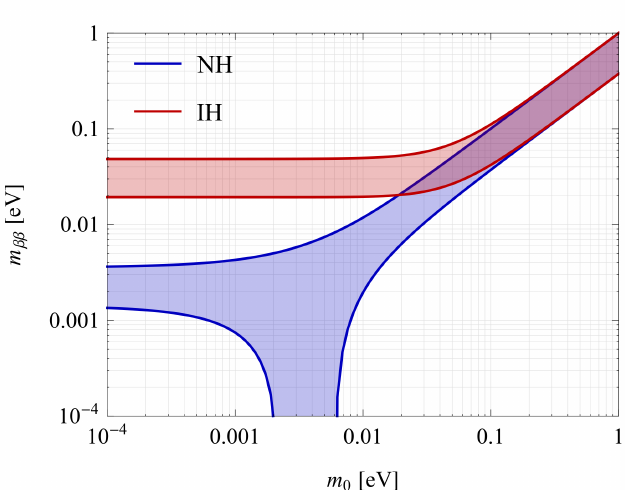}
\caption{Scenario \ref{sec:Uncoupled} with 
$m_{i} M_{i} = const$.  The effective Majorana neutrino mass $m_{\beta \beta}$
as a function of the lightest neutrino mass for 
the normal (blue) and inverted (red) hierarchy of neutrino
masses. The best fit values of neutrino oscillation parameters
from the global analysis of neutrino oscillation
data \cite{globfit16} are considered. 
%If there is no coupling
%between light and heavy neutrino sectors and $V_{0}=U$ (U is the
%PMNS mixing matrix), for $\lambda \, \langle p^2 \rangle_a /\zeta_p =1$
%we have $M^R_{\beta \beta} = m_{\beta\beta}$.}
}
\label{fig:lnv01}
\end{figure}

Let us give a couple of examples of model inputs allowing us to distinguish 
two above-mentioned mechanisms. 

%There is the possibility of discussing the general lepton number violating
%parameter $\eta_{\nu N}$ for light and heavy neutrino mass mechanisms
%within some viable particle physics scenarios. We shall address some
%of them.
%
For a scenario with three SM singlet neutrinos $\nu_{e, \mu, \tau R}$ 
the $6\times 6$ mixing matrix $\mathcal{U}$ in Eq. (\ref{maticazmies}) is completely parameterized
with 15 angles, 10 Dirac and 5 Majorana CP violating phases. Let
consider some viable structures of this mixing matrix.

\begin{figure}[t]
\centering
\includegraphics[width=0.5\textwidth]{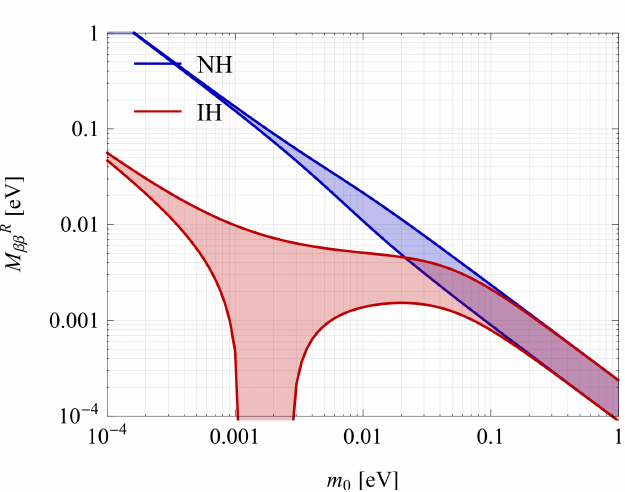}
\caption{Scenario \ref{sec:Uncoupled} with 
$m_{i}/M_{i} = const$. 
The effective Majorana neutrino mass $M^R_{\beta \beta}$ 
as a function of the lightest neutrino mass $m_{0}$ for 
the normal (blue) and inverted (red) hierarchy of neutrino
masses. 
%The uncoupled light and heavy neutrino sectors are considered
%and $V_{0}=U$. The relation between light and heavy neutrino masses is
%$M_i = \zeta_p/m_i$ (constant products) with $\zeta_r = 10^{-17}$.
%For squared ratio of masses of left and right vector bosons
%it is assumed $\lambda = 7.7 \cdot 10^{-4}$.}
}
\label{fig:lnv02}
\end{figure}

\subsection{Uncoupled light and  heavy neutrino sectors}
\label{sec:Uncoupled}
In the particular case of
\begin{eqnarray}
 {\mathcal{U}} &=& \left(
\begin{array}{ll}
U_0 & \mathbf{0}\\
 \mathbf{0} & V_0\\
 \end{array}
\right).
\label{maticaXing}
\end{eqnarray}
there is no mixing between heavy and light neutrino sectors. Then we have
\begin{equation}
\label{eq:eta}
\eta_{\nu N}^2 = \frac{1}{m_e^2} \left( m_{\beta \beta}^2 + (M^R_{\beta \beta})^2 \right)
\end{equation}
\begin{eqnarray}
\label{eq:mbb}
m_{\beta \beta}
&=& \left| \sum_{j=1}^3 U_{ej}^2 \, \frac{m_j}{1 + m_j^2/\langle p^2 \rangle_a} \right| 
\simeq \left| \sum_{j=1}^3 U_{ej}^2 \, m_j \right|, \\
\label{eq:mmbb}
M^R_{\beta \beta}
&=& \lambda \left| \sum_{k=1}^3 U_{ek}^2 \, \frac{M_k}{1 + M_k^2/\langle p^2 \rangle_a} \right| 
\simeq \lambda \, \langle p^2 \rangle_a \left| \sum_{k=1}^3 U_{ek}^2 \, \frac{1}{M_k} \right|.\nonumber\\
\end{eqnarray}
In this scenario $U_0$ can be identified with the PMNS mixing matrix $U$. 
Thus we assume $U_{0}=U$.  
The mixing matrix $V_0$ for the heavy neutrinos is unknown, but it is similar to $U_{0}$ in the light neutrino sector, then $V_{0}=U$ is frequently assumed.
For sake of simplicity we
%assume 
consider
two different cases for the heavy neutrino masses:
\begin{equation}
\label{eq:mm}
M_i =
\begin{cases}
m_i/\zeta_r & \textrm{constant ratios}\\
\zeta_p/m_i & \textrm{constant products}
\end{cases}
\end{equation}
In the case of 
%uncoupled sectors (\ref{eq:mmbb}) and 
the constant products $\zeta_p = m_i M_i$ we have for the LNV parameter in Eq.~(\ref{eq:eta}):
%(\ref{eq:mm}) is proportional to $m_{\beta \beta}$ multiplied by a factor of
%$\lambda \, \langle p^2 \rangle_a /\zeta_p$, 
\begin{eqnarray}\label{eq:Uncoupl-ConstProd-1}
  \eta_{\nu N}^2 = \frac{1}{m_e^2} \left(1 + \lambda^{2} \, \left(\frac{\langle p^2 \rangle_a }{\zeta_p}\right)^{2} \right)
  \, m_{\beta \beta}^2 \equiv \kappa^{2} \, m_{\beta \beta}^2.\nonumber\\
%M^{R}_{\beta \beta} &=& \lambda \, \frac{\langle p^2 \rangle_a }{\zeta_p}\,  m_{\beta \beta}
%
\end{eqnarray}
Thus, in this scenario the presence of heavy neutrinos leads to a vertical shift 
%which results only in a vertical shift 
of the standard plot in Fig.~\ref{fig:lnv01} by a constant factor 
$\kappa$.   As a result, the $0\nu\beta\beta$-decay experimental upper bound on $m_{\beta\beta}$ is significantly
less stringent, if $\zeta_p \ll \lambda \langle p^2 \rangle_a \simeq $ 24 MeV$^2$.
%
%
%The plot is given for 
%the best fit values of the neutrino oscillation parameters derived from the global analysis of the neutrino oscillation data \cite{globfit16}.
%the effective Majorana neutrino mass $m_{\beta \beta}$ (\ref{eq:mbb}) is shown in Fig.\ \ref{fig:lnv01}.
%
%The right-handed current coupling constant $\lambda = \left( M_{W_L}/M_{W_R} \right)^2$ is
%currently constrained by the limit: $M_{W_R} > 2.9 \, \mathrm{TeV}$.  
%In what follows, 
%we assume the most optimistic value: 
In our estimation we used the upper-bound-value in (\ref{eq:lambda-limit}), i. e.
$\lambda = 7.7 \cdot 10^{-4}$, and $\sqrt{\langle p^2\rangle_a}$ = 175 MeV
calculated within the QRPA  by assuming Argonne potential and $g_A = 1.27$ (see Table~\ref{table.2}).
% yields an average Majorana-neutrino momentum:
%$\sqrt{\langle p^2 \rangle_a} = 175 \, \mathrm{MeV}$ (irrespective of the considered isotope).
%The heaviest among the light neutrinos must possess mass of at least $50 \, \mathrm{meV}$
%(given by the large mass gap $\sqrt{|\Delta m^2|}$), which justifies the expected light-neutrino
%mass scale: $m_i \sim 10^{-1} \, \mathrm{eV}$.

In the case of 
%uncoupled sectors (\ref{eq:mmbb}) and 
the constant ratios $\zeta_r = m_i/M_i$ in Eq.~(\ref{eq:mm})
the effective Majorana neutrino mass $M_{\beta \beta}$ is shown in Fig.\ \ref{fig:lnv02}.
Contribution of $M_{\beta \beta}$ becomes comparable to $m_{\beta \beta}$ as soon as $\zeta_r = 10^{-17}$,
which corresponds to $M_i \sim 10^{16} \, \mathrm{eV} = 10^4 \, \mathrm{TeV}$. $\lambda = 7.7 \cdot 10^{-4}$
is assumed again. Notice the reversed
behavior of the mass hierarchies: NH no longer exhibits a region unbounded from below, while IH does.

\begin{figure}[t]
\centering
\includegraphics[width=0.5\textwidth]{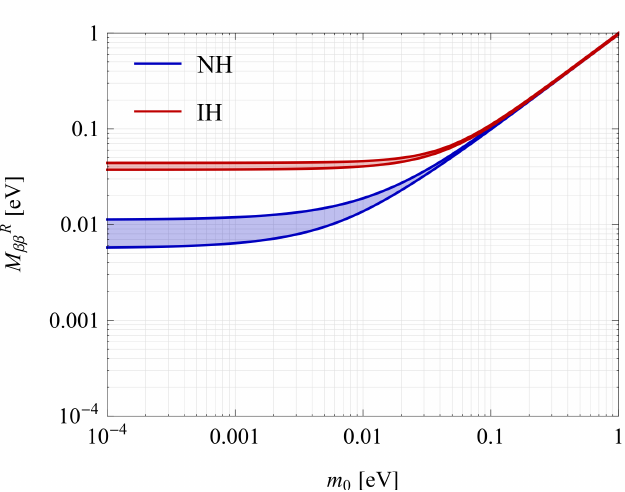}
\caption{Scenario \ref{sec:SeesawLike}.
  The effective Majorana neutrino mass $M^R_{\beta \beta}$  in Eq.~(\ref{eq:MR-1}) with $m_{D}=5$ MeV
  as a function of the lightest neutrino mass $m_{0}$ for  
the normal (blue) and inverted (red) hierarchy of neutrino
masses. 
%The coupled light and heavy neutrino sectors are considered
%with  $V_{0}=U^\dagger$.
%The relation between light and heavy neutrino masses is
%$m_i \simeq m^2_D/M_{i}$ with $m_{\mathrm{D}} \approx 5 \, \mathrm{MeV}$.
%For squared ratio of masses of left and right vector bosons
%it is assumed $\lambda = 7.7 \cdot 10^{-4}$.}
}
\label{fig:lnv04}
\end{figure}

\subsection{Seesaw-mixed light and heavy neutrino sectors}
\label{sec:SeesawLike}
Assuming for simplicity the flavor universal mixing between the active and sterile neutrino sectors
the seesaw mixing matrix $\mathcal{U}$ takes the form
\begin{eqnarray}
\mathcal{U}&=& \left(
\begin{array}{cc}
U_0 & \zeta~ \mathbf{1} \\
- \zeta~\mathbf{1} & V_0 \\
 \end{array}
\right). 
\label{specmixing}
\end{eqnarray}
Here, $\zeta=\frac{m_D}{m_{LNV}}$, where $m_D$ is the typical scale of  the charged leptons masses and $m_{LNV}$ is the LNV scale of the order of the Majorana masses $M_{i}$  of the heavy neutrinos.  As in the previous scenario $U_0$ can be identified with the PMNS $U$ matrix. Thus we assume $U=U_0$.  For $V_0$,  analogue of $U_{0}$ in the heavy neutrino sector, we find from the unitarity conditions\\[-15mm]
\begin{equation} 
 V_0 = U^\dagger_0
\end{equation}
\mbox{}\\[-15mm]
and\\[-15mm]
\begin{equation} 
 U_0 U^\dagger_0 = (1-\zeta^2)\mathbf{1},~~~~V_0 V^\dagger_0 = (1-\zeta^2)\mathbf{1}.
\end{equation}
\mbox{}\\[-15mm]
It is assumed that a small violation of the unitarity of $U_0$ and $V_0$ matrices is beyond
the current accuracy of phenomenological determination of elements of the PMNS matrix. 
The matrix $V_{0}$ takes the form
\begin{widetext}
\begin{equation}
V_{0} = U^{\dagger}_0 =
\begin{pmatrix}
c_{12} \, c_{13} \, e^{-i \alpha_1} & \left( -s_{12} \, c_{23} - c_{12} \, s_{13} \, s_{23} \, e^{-i \delta} \right) e^{-i \alpha_1} & \left( s_{12} \, s_{23} - c_{12} \, s_{13} \, c_{23} \, e^{-i \delta} \right) e^{-i \alpha_1} \\
s_{12} \, c_{13} \, e^{-i \alpha_2} & \left( c_{12} \, c_{23} - s_{12} \, s_{13} \, s_{23} \, e^{-i \delta} \right) e^{-i \alpha_2} & \left( -c_{12} \, s_{23} - s_{12} \, s_{13} \, c_{23} \, e^{-i \delta} \right) e^{-i \alpha_2} \\
s_{13}\, e^{i \delta} & c_{13} \, s_{23} & c_{13} \, c_{23} 
\end{pmatrix}.
\end{equation}
\end{widetext}
We note that each element of the first row is multiplied by the same phase factor $e^{-i \alpha_1}$. Analogously, the second raw is multiplied by  $e^{-i \alpha_2}$. Therefore,  the
Majorana phases $\alpha_{1,2}$ do not affect the heavy neutrino LNV parameter 
$M^R_{\beta\beta}$ in this case. On the contrary, the Dirac phase $\delta$, which does not affect the light neutrino LNV parameter  $m_{\beta\beta}$ 
will impact the value of $M^R_{\beta\beta}$. 
The seesaw structure of (\ref{specmixing}) implies $m_i \simeq m^2_D/m_{LNV}$ 
and $M_i  \simeq m_{LNV}$.
For a product of light and heavy neutrino masses let assume $m_i M_i \simeq m^2_{D}$.
If the LNV scale is significantly larger than $\langle p^2\rangle_a$ we find 
\begin{eqnarray}
\label{eq:EtaNu-1}
  \eta_{\nu N}^2 &=& \frac{1}{m^2_e}
  \left(m^2_{\beta\beta} ~+~ (M^{\rm R}_{\beta\beta})^2\right)
\end{eqnarray}
with
\begin{eqnarray}
\label{eq:MR-1}
M^{\rm R}_{\beta\beta} = \lambda ~\frac{\langle p^2\rangle_a}{m^2_D}~
\left|\sum_{j=1}^3(U^\dagger_0)^{2}_{ej}~ m_j \right|.
\end{eqnarray}
We note that for $m_D \simeq 5$~MeV 
the coefficient $\lambda ~{\langle p^2\rangle_a}/{m^2_D}$ enetring 
$M^R_{\beta\beta}$ in Eq. (\ref{eq:MR-1}) is close to unity and it might
be that contributions from the light and heavy neutrinos to $\eta_{\nu N}$
are comparable. However, $M_{\beta\beta}^R$ is not proportional to $m_{\beta\beta}$ as
off-diagonal elements of matrices $U_{0}$ and $(U^\dagger)$ are different.
Therefore, a detailed analysis is needed to establish an useful 
constraint on the Yukawa potential associated with neutrinos.
In Fig. \ref{fig:lnv04} we show $M_{\beta\beta}^R$ as function of lightest neutrino mass
both for normal and inverted hierarchy by assuming $m_{\mathrm{D}} \simeq 5 \, \mathrm{MeV}$ 
(and $\lambda = 7.7 \cdot 10^{-4}$). 

\begin{figure}[t]
\centering
\includegraphics[width=0.5\textwidth]{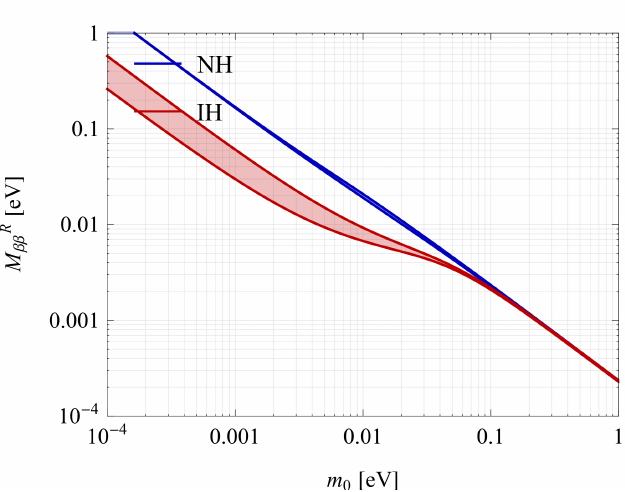}
\caption{The same as in Fig.~\ref{fig:lnv04}, but for  
$M^R_{\beta \beta}$ defined according to Eq.~(\ref{eq:MR-2}) with 
$\zeta^2 = 10^{-17}$.
%as a function of the lightest neutrino mass for cases of 
%the normal (in blue) and inverted (in red) hierarchy of neutrino
%masses. The coupled light and heavy neutrino sectors are considered
%with  $V_{0}=U^\dagger$.
%The relation between light and heavy neutrino masses is
%$\zeta^2 = m_i/M_i$ with $\zeta^2 = 10^{-17}$.
%For squared ratio of masses of left and right vector bosons
%it is assumed $\lambda = 7.7 \cdot 10^{-4}$.
}
\label{fig:lnv03}
\end{figure}

Within the seesaw structure one can also assume $m_i \simeq \zeta^2 M_i$. Then
we find
\begin{eqnarray}
\label{eq:MR-2}
M^{\rm R}_{\beta\beta}
  = \lambda~\zeta^2~
\left|\sum_{j=1}^3(U^\dagger_0)^{2}_{ej}~ \frac{\langle p^2\rangle_a}{m_j}\right|.
\end{eqnarray}
For $\zeta^2 = 10^{-17}$ and $\lambda = 7.7 \cdot 10^{-4}$
the effective mass $M^{\rm R}_{\beta\beta}$ in Eq. (\ref{eq:MR-2}) is plotted
in Fig.~\ref{fig:lnv03}. 
We see again that for a chosen set of parameters the value of $M^{\rm R}_{\beta\beta}$ can be comparable with $m_{\beta\beta}$
(see Fig. \ref{fig:lnv01}).

\begin{figure}[t]
\centering
\includegraphics[width=0.45\textwidth]{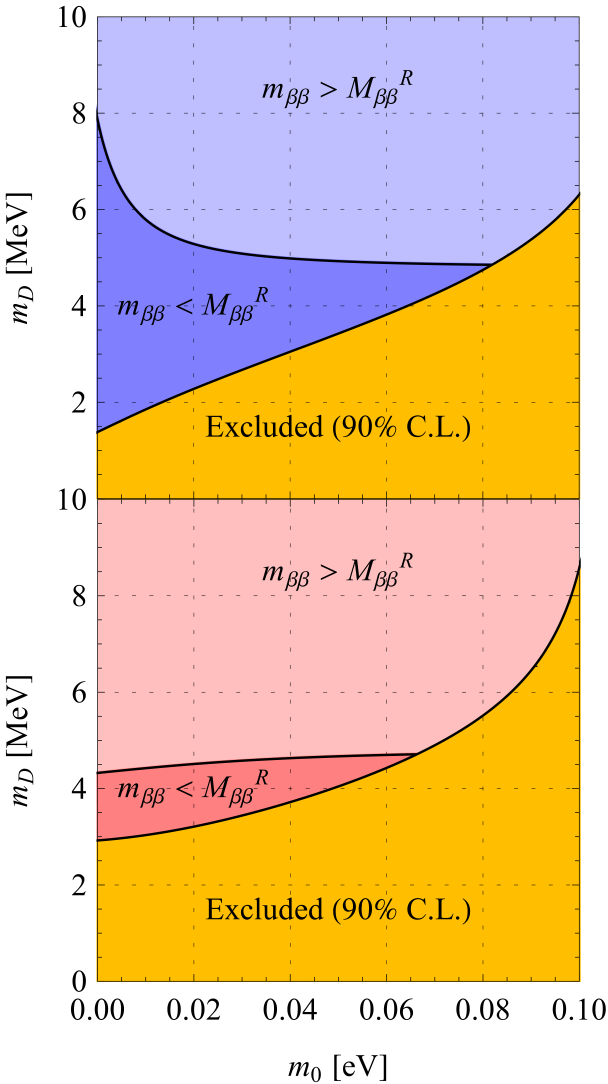}
\caption{Scenario \ref{sec:SeesawLike} with the mass relation \mbox{$m_i \simeq m^2_D/M_{i}$.}
The comparison of  the light $m_{\beta\beta}$ and heavy $M^R_{\beta\beta}$ neutrino 
contributions to $0\nu\beta\beta$-decay for the normal(inverted) hierarchy is shown in left(right) panel.  
%  of neutrino masses are  presented in the left and right pannels, respectively. 
%  over $m_{\beta\beta}$ contribution to the LNV parameter $\eta_{\nu N}$
%  for the see-saw type of neutrino mixing matrix given in Eq. (\ref{specmixing}) 
%  and by assuming $m_i \simeq m^2_D/M_{i}$ The cases of the normal and inverted hierarchy
%  of neutrino masses are  presented in the left and right pannels, respectively.
%For squared ratio of masses of left and right vector bosons
%$\lambda = 7.7 \cdot 10^{-4}$ is considered.
%  The current constraint on the
%$\eta_{\nu N}$ parameter deduced from the lower limit on the $0\nu\beta\beta$-decay
%half-life of $^{136}$Xe \cite{expXe} with help of QRPA (TBC) NME \cite{TBC13}
%is considered.
%  For squared ratio of masses of left and right vector bosons
%$\lambda = 7.7 \cdot 10^{-4}$ is assumed. 
}
\label{fig:lnv05}
\end{figure}

In Table \ref{table.limit} we show 
%Taking into account 
%For the analysis of the dominance of the light or heavy neutrino exchange
%contributions to the LNV parameter 
%
upper bounds on $\eta_{\nu N}$ in Eq.~(\ref{eq:EtaNu-1}) derived from the current  $0\nu\beta\beta$-decay experiments.  
%
%have to be taken into account.
%the current experimental limits on $0\nu\beta\beta$-decay half-life
%are displayed in Table \ref{table.limit}.  The corresponding 
%upper bounds on the parameter $\eta_{\nu N}$ are deduced from
%by using the coefficients $C_{\nu N}$  of Table \ref{table.limit}
%assuming an unquenched value of the axial-vector coupling constant.
As seen, the most stringent  bound comes out from  $^{136}$Xe $0\nu\beta\beta$-decay experiment Ref.~\cite{expXe}.  
%and $^{76}$Ge set the sharpest limit
%$\eta_{\nu N}\le$ $2.9\times 10^{-7}$ and $1.2\times 10^{-7}$, 
%respectively. 
For this bound we analyzed the separate contributions of the light and heavy neutrinos to $0\nu\beta\beta$-decay. 
Fig.~\ref{fig:lnv05} displays the corresponding results in the plane of the parameters
$m_D$ and $m_0$ ($m_i \simeq m^2_D/M_{i}$ is assumed) for the cases of 
%the region of dominance of $M^R_{\beta\beta}$ mechanism
%over $m_{\beta\beta}$  mechanism in the $0\nu\beta\beta$-decay
%rate is displayed. The cases of 
the normal (left panel)
and inverted hierarchy (right panel) of neutrino masses.
%are discussed. The current limit on the half-life of $^{136}$Xe
%\cite{expXe} and  QRPA (TBC) NMEs with Argonne short-range correlations
%\cite{TBC13} are taken into account. The unquenched value of
%the axial-vector coupling constant is assumed ($g_A=1.27$). For squared
%ratio of masses of left and right vector bosons it is assumed
%$\lambda = 7.7 \cdot 10^{-4}$. 
%By glancing
%the 
We see that in the considered scenario
%we see that in the case of 
for normal (inverted)
hierarchy 
the values $m_D \le 1.5$ MeV ($m_D \le 2.9$ MeV) are already excluded by
the existing experimental data on $0\nu\beta\beta$-decay.
%:{SK_21.02.2018-1} Fedia, ya ne ponial etu frazu
%By keeping in mind this restriction we can conclude that the mass of the
%lightest is larger about 30 MeV. 
%:{SK_21.02.2018-1} END
We also see that in the case of normal (inverted) neutrino mass hierarchy
the heavy neutrino exchange mechanism cannot dominate over the light 
one in the region \mbox{$m_0 \ge 0.08$ eV} ($m_0 \ge 0.065$ eV).
The contraint from the $0\nu\beta\beta$-decay experiment imply that 
the limit on the  mass of lightest heavy neutrino is $M_3 >$ 38 TeV and
$M_2 >$ 171 TeV in the cases of normal and inverted hierarchy, respectively.

\begin{figure}[t]
\centering
\includegraphics[width=0.45\textwidth]{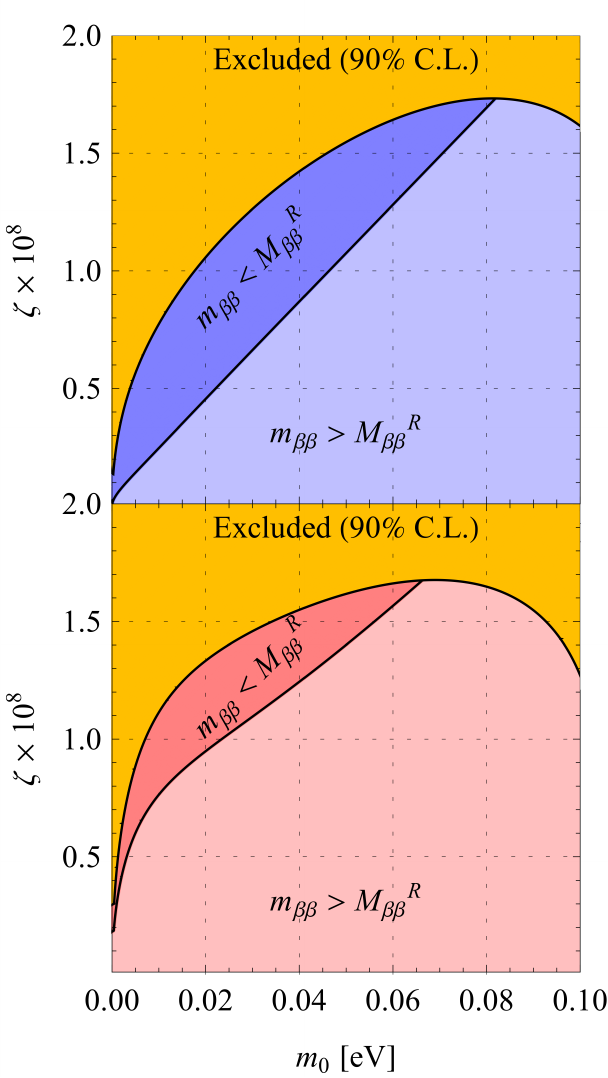}
\caption{The same as in Fig.~\ref{fig:lnv05}, but 
with the mass relation $\zeta^2 = m_i/M_i$.
%  The region of the dominance of the $M^R_{\beta\beta}$ contribution 
%  over $m_{\beta\beta}$ contribution to the LNV parameter $\eta_{\nu N}$
%  for the see-saw type of neutrino mixing matrix given in Eq. (\ref{specmixing}) 
%  and by assuming $\zeta^2 = m_i/M_i$. The cases of the normal and inverted hierarchy
%  of neutrino masses are  presented in the left and right pannels, respectively.
%  The current constraint on the
%$\eta_{\nu N}$ parameter deduced from the lower limit on the $0\nu\beta\beta$-decay
%half-life of $^{136}$Xe \cite{expXe} with help of QRPA (TBC) NME \cite{TBC13}
%is considered.
%  For squared ratio of masses of left and right vector bosons
%$\lambda = 7.7 \cdot 10^{-4}$ is assumed. 
}
\label{fig:lnv06}
\end{figure}

Fig. \ref{fig:lnv06} shows results of an analysis similar to the above-discussed one,
%that presented in 
%Fig. \ref{fig:lnv05} is shown 
but for $\zeta^2 = m_i/M_i$ scenario.
%
%the same nuclear systems, experimental lower bound on the
%half-life, NME and value of $\lambda$. 
%
In this case the heavy neutrino mechanism cannot dominate in practically the same domain of $m_{0}$
as previously.
%
%Practically,
%the same restrictions in respect of $m_0$ 
%concerning the dominance of $M^R_{\beta\beta}$ mechanism remains valid.
%:{SK_21.02.2018-3} Fedia, ya ne ponial etu frazu
It is concluded that in the case of normal (inverted)
hierarchy $\zeta \le 1.75\times 10^{-8}$  ($\zeta \le 1.65\times 10^{-8} $).
%:{SK_21.02.2018-3} END
We note that within the considered seesaw scenario within the LRSM the effective Majorana neutrino mass $m_{\beta\beta}$
can not be identified with the first element of the Dirac-Majorana mass (see Appendix \ref{app:B}) $(M_L)_{ee}$, which
contains additional term $\zeta^2 M_1$ in magnitude comparable with $m_1$. The corresponding term in $m_{\beta\beta}$
has been neglected as it is suppressed by properties of neutrino propagator for large neutrino mass. Due to the same
reason $M_{\beta\beta}$ can not be identified with $(M_R)_{ee}$.

\begin{table*}[!t]
  \begin{center}
    \caption{
      Upper bounds on the effective lepton number violating 
      parameter $\eta_{\nu N}$ imposed by the current constraints on the $0\nu\beta\beta$-decay
      half-life $T^{0\nu-exp}_{1/2}$ (the first row). The values in the second and the third rows
      were obtained using the largest and lowest values of $C_{\nu N}$ 
      for a given isotope from  Table \ref{table.2}, respectively. 
%      The unquenched
%      value of the axial-vector coupling constant $g_A$ is assumed. 
      \label{table.limit}}
 \begin{tabular}{lcccccccc}\hline\hline
 & & $^{48}$Ca & $^{76}$Ge & $^{82}$Se & $^{100}$Mo & $^{116}$Cd  & $^{130}$Te &
   $^{136}$Xe   \\ \hline
   $T^{0\nu-exp}_{1/2}$ [yrs]   & &
   $2.0\times 10^{22}$ \cite{expCa} &    $5.3\times 10^{25}$ \cite{expGe} &  $2.5\times 10^{23}$ \cite{expSe}
   &  $1.1\times 10^{24}$ \cite{expMo}  &    
   $1.7\times 10^{23}$ \cite{expCd} &    $4.0\times 10^{24}$ \cite{expTe} &   $1.07\times 10^{26}$  \cite{expXe}   \\
   $\eta_{\nu N}\times 10^{6}$      & 
   & 10.3 &   0.290 & 2.32 & 0.724 &  2.14 &  0.532 & 0.117  \\                 
   &   & 33.8 &  0.643  & 4.81 & 1.22  &  3.76 &  1.455   & 0.306  \\                 
   \hline\hline   
      \end{tabular}
  \end{center}
\end{table*}

%%%%%%%%%%%%%%%%%%%%%%%%%%%%%%%%%%%%%%%%%%%%%%%%%%%%%%%%%%%%%%%%%%%%%%%%%%%%
%%%%%%%%%%%%%%%%%%%%%%%%%%%%%%%%%%%%%%%%%%%%%%%%%%%%%%%%%%%%%%%%%%%%%%%%%%%%
\section{\label{sec:conclusions}Conclusions}
%%%%%%%%%%%%%%%%%%%%%%%%%%%%%%%%%%%%%%%%%%%%%%%%%%%%%%%%%%%%%%%%%%%%%%%%%%%%
%%%%%%%%%%%%%%%%%%%%%%%%%%%%%%%%%%%%%%%%%%%%%%%%%%%%%%%%%%%%%%%%%%%%%%%%%%%%

In summary, we have shown that the ratio of nuclear matrix elements for the light
and heavy neutrino mass mechanisms exhibits practically no dependence on
isotope for all favoured nuclear structure methods. This quantity, when
properly scaled, can be identified with squared average neutrino momentum $\langle p^2\rangle$
of the interpolating formula including light and heavy neutrino exchange mechanisms.
%within left-right symmetric models. 
The universality of the averaged value of
$\langle p^2\rangle$ for a set of isotopes allows determination of  a new 
%lepton number 
LNV parameter
$\eta_{\nu N}$,  which is a coherent sum of squared 
%lepton number violating 
LNV
parameters
$m_{\beta\beta}$ and $M^R_{\beta\beta}$ characterizing the light and heavy neutrino exchange mechanisms,
respectively. Thus, the observation of $0\nu\beta\beta$-decay on two and
more nuclear isotopes will allow one to deduce information about  the size of $\eta_{\nu N}$, but not 
about the relative contribution of light or heavy neutrino-exchange  mechanism
to the decay rate. 
An additional theoretical or experimental input 
%or model assumptions concerning 
about neutrino masses and mixing is
needed to shed light on the particular role of each of these mechanisms. 
%The dominance of any of these mechanisms can be discussed by making assumption  
%concerning the mixing of light and heavy neutrinos and the relation
%among their masses. 
As an example we considered a simplified see-saw type
$6\times 6$ neutrino mixing matrix (\ref{specmixing}), which implies
that the $3\times 3$ mixing matrix of heavy neutrinos is the hermitian conjugate
of the $3\times 3$ PMNS mixing matrix of light neutrinos. Assuming
several viable seesaw relations among the light $m_i$ and heavy $M_i$ neutrino masses
(i=1,2 and 3) useful constraints on the parameters, in particular Dirac neutrino mass $m_D$,  entering these relations have been obtained from  the experimental lower bounds on the
$0\nu\beta\beta$-decay half-life
% of $^{136}$Xe. 
The region of dominance of heavy over light neutrino exchange mechanisms for the considered scenarios have been identified.

\begin{acknowledgments}
This work is supported by the VEGA Grant Agency of the Slovak Republic un- der Contract No. 1/0922/16,
by Slovak Research and Development Agency under Contract No. APVV-14-0524, RFBR Grants
Nos. 16-02-01104 and 18-02-00733, Underground laboratory LSM - Czech participation to European-level research
infrastructure CZ.02.1.01/0.0/0.0/16\_013/0001733,  Fondecyt (Chile) grant 
 No.~1150792,  and by CONICYT (Chile) Ring ACT1406, PIA/Basal FB0821.
\end{acknowledgments}

\appendix 

\section{Analytical Properties of the NMEs and the Interpolating formula}
\label{app:A}
Here we give some comments on the possible improvement of our interpolating formula in 
Eq.~(\ref{eq:interpol-form-1}), which we call the ``monopole'' approximation. 
Numerically the latter is already a very good approximation to the ``exact'' NMEs given by 
Eq.~(\ref{eq:MnuN}) and calculated in the framework of any specific nuclear structure approach. However in certain cases one may need an approximate formula having not only a good numerical precision, but also the analytical properties in the complex plane of $m_{\nu}$ the same or maximally close to the ``exact'' NME defined in expression (\ref{eq:MnuN}). 

Obviously the monopole approximation (\ref{eq:interpol-form-1})
has two imaginary poles in the complex plane of $m_{\nu}$, while they are absent in
the exact expression (\ref{eq:MnuN}).
%A more accurate interpolation could take place using a function whose analytic properties
%coincide with those of the exact expression (\ref{eq:MnuN}).
%We describe a class of such functions.
Below we describe a class of  approximations with the analytic properties
of the ``exact'' NME (\ref{eq:MnuN}).
%We describe a class of such functions.

%Eq.~(\ref{eq:MnuN}) defines an analytic function of the neutrino mass.
%It is known from complex analysis that behavior of analytic function in a fixed domain
%is determined by the closest singularities. 
%Equation (\ref{eq:interpol-form-1})
%has two imaginary poles with respect to neutrino mass,
%however, these poles are not in the exact expression (\ref{eq:MnuN}).
%A more accurate interpolation could take place using a function whose analytic properties
%coincide with those of the exact expression (\ref{eq:MnuN}).
%We describe a class of such functions.

Let us rewrite Eq.~(\ref{eq:MnuN}) 
in the form
%can be written as follows
\begin{equation}
{M}_{LL,RR}^{\prime \,0\nu }(m_{\nu}) =
\frac{4\pi}{(2\pi )^{3}}  \int_0^{\infty} p^2 d{p}
\frac{ \varphi({p})}{E_p(E_p+\Delta )},
\label{AN2}
\end{equation}%
where $\Delta =E_{n}-(E_{I}-E_{F})/2>0$, $E_p=\sqrt{p^{2}+m_{\nu}^{2}}$,
\begin{equation}
\varphi({p}) =
\int d\mathbf{x}d\mathbf{y} e^{i\mathbf{p\cdot }(\mathbf{x}-\mathbf{y})}
\varphi(\mathbf{x},\mathbf{y}),
\label{ANp}
\end{equation}%
and
\begin{equation}
\varphi(\mathbf{x},\mathbf{y}) = \frac{1}{m_{\mathrm{p}}m_{\mathrm{e}}}\frac{4\pi R}{g_{A}^{2}}\sum_{n}
\langle 0_{F}^{+}|J_{LR}^{\mu \dag }(\mathbf{x}%
)|n\rangle \langle n|J_{\mu LR}^{\dag }(\mathbf{y})|0_{i}^{+}\rangle. \label{BP}
\end{equation}%
The function $\varphi(\mathbf{x},\mathbf{y})$ describes a distribution of currents inside the nucleus.
In Eq.~(\ref{AN2}) the neutrino mass enters the denominator of the integrand.

Analytic properties of functions defined in terms of a contour integral are fixed by the Landau rules \cite{Landau:original,Landau}.

The singular points of the first kind are associated with singular behavior of the integrand at the end points of the integration contour.
In the case of Eq.~(\ref{AN2}), these singularities could occur provided that 
\mbox{$\chi(p) \equiv E_p(E_p+\Delta ) = 0$} for $p=0$ or $\infty$.
This equation can be fulfilled for $p=0$ only to give $ m = 0$ and $m = \pm \Delta $.
The points $m = \pm \Delta $ are located on the different sheets of the Riemann surface
of ${M}_{LL,RR}^{\prime \,0\nu }(m_{\nu})$.
It is clear that model dependent features of the nuclear structure entering $\varphi(\mathbf{x},\mathbf{y})$ do not affect the end point singularities.

Singular points of the second kind are associated with the pinch singularities of the integrand.
To find them, the equations $\chi(p)/\varphi(p)  = 0$ and $(\chi(p)/\varphi(p))^{\prime}  = 0$ are to be solved,
which localise high-order poles of the integrand in the complex $ p $--plane.
These singularities depend on $\varphi(\mathbf{x},\mathbf{y})$ and thereby on the nuclear structure model.

Analytic properties of ${M}_{LL,RR}^{\prime \,0\nu }(m_{\nu})$ as a function of $\Delta$ are particularly simple.
Changing the variable in Eq.~(\ref{AN3}) to $p=m\sinh\theta$,
we arrive at the dispersion integral
\begin{equation}
{M}_{LL,RR}^{\prime \,0\nu }(m_{\nu})=\frac{4\pi m}{(2\pi )^{3}}\int_{0}^{\infty
}\sinh ^{2}\theta d\theta \frac{\varphi (m\sinh \theta )}{\cosh \theta
- \xi},  \label{AN3}
\end{equation}
where $\xi = - \Delta/m_{\nu}$. This equation shows that
${M}_{LL,RR}^{\prime \,0\nu }$ is an analytic function in the complex $\xi$--plane with the cut $(1,+\infty)$ corresponding to the cut $(-\Delta,0)$ in $m_{\nu}$.
%The discontinuity in $m$ of ${M}_{LL,RR}^{\prime \,0\nu }(m)$ across the cut $(-\Delta,0)$ can also be found.
%In the case of $\varphi (p)$ is analytic function for $|p| < \infty$ and the integral (\ref{AN3}) converges,
Provided $\varphi (p)$ is an analytic function for $|p| < \infty$ and the integral (\ref{AN3}) converges,  ${M}_{LL,RR}^{\prime \,0\nu }(m)$ turns out to be an analytic function in the complex $m_{\nu}$--plane with the cut $(-\Delta,0)$.
On the second sheet of the Riemann surface one finds a branch point $m = + \Delta$.

\begin{figure} [t] %
\begin{center}
\includegraphics[width=0.5\textwidth]{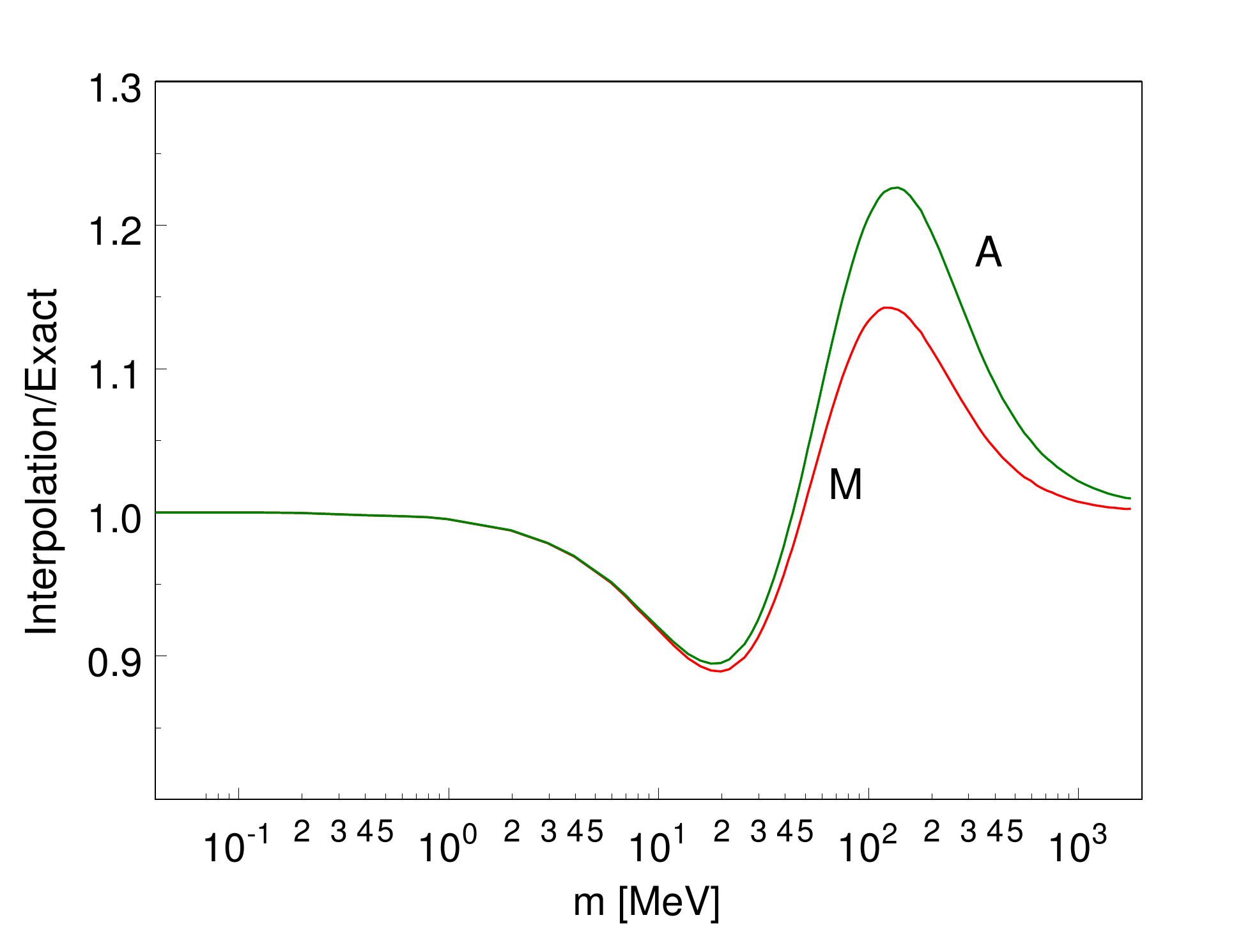}
%\includegraphics[angle = 0,width=0.49\textwidth]{GS5.eps}
%\vspace{-10cm}
\caption{(color online)
Ratio between the interpolation formulas and the exact calculation for $^{76}$Ge versus neutrino mass. 
The curve M is for the monopole interpolation (\ref{eq:interpol-form-1}),
the curve A shows the ratio for the analytic interpolation formula.
}
\label{Fig15}
\end{center}
\end{figure}

As we discussed before the monopole parametrization 
(\ref{eq:interpol-form-1}) is 
numerically very accurate. This parametrization corresponds to an approximation of the spectral function with the delta function:
$ \phi_{\mathrm{m}}(p) \sim \delta (p^{2} - \langle p^2\rangle)$.
%In order for the analytic formula to work  successfully, 
Then for the formula with the correct analytical properties, which we are going to construct here, 
we chose the spectral function in a form close to the $ \phi_{\mathrm{m}}(p)$ to guarantee its numerical accuracy comparable with the monopole parameterization. We may choose
\begin{equation} \label{spectral fun}
\phi(p) = \exp\left(- \frac{\rho^2 p^2}{2}\right)\frac{\sinh(\rho^2 p p_0 )}{\rho^2 p p_0}.
\end{equation}
with the free parameters, $\rho$ and $p_0 \sim \langle p^2\rangle^{1/2}$, which can be fixed by normalization to the exact values at zero and at infinity.
The function $\phi(p)$ for $p = p_0$ is close to the maximum, the value of $1/\rho$ determines the width of momentum distribution.
This spectral function is analytic for $|p| < \infty$ and it generates model independent end-point singularities only.
The corresponding interpolating formula appears to be an analytic function in the complex $m_{\nu}$--plane with the cut $(-\Delta,0)$.
The cut position is model-independent. The discontinuity depends on $\phi(p)$ and is model dependent.
A particularly strong effect on the behavior of analytic functions in a fixed domain comes from nearest singularities.
Taking into account that $\Delta \sim 10$ MeV, an improved description of the neutrino mass dependence can be expected
around zero neutrino mass in the circle with a radius of a few tens of MeV. This scale is smaller than 
the characteristic momentum transfer $p_0 \sim 200$ MeV.
Reasonable accuracy  is also expected  for large $m_{\nu}$ domain, provided the spectral function (\ref{spectral fun}) 
approximates closely the monopole spectral function found to be successful phenomenologically.\\
\mbox{\hspace{3mm}}The ratio between the interpolating formula of Eq.~(\ref{eq:interpol-form-1}) and the exact calculation for $^{76}$Ge 
is shown on Fig.~\ref{Fig15}. The result is compared to the interpolating formula of the spectral function (\ref{spectral fun}) 
with $\rho = 5$ fm and $p_0 = 0.84$/fm. For low neutrino masses up to about 40 MeV
the analytic interpolation formula approximates the exact result with a better accuracy. 
For higher masses the nuclear structure at about 200 MeV
becomes important, which could reflect a contribution of the model dependent pinch singularities which we do not consider. 

%%%%%%%%%%%%%%%%%%%%%%%%%%%%%%%%%%%%%%%%%%%%%%%%%%%%%%%%%%%%%%%%%%%%%%%%%%%%
%%%%%%%%%%%%%%%%%%%%%%%%%%%%%%%%%%%%%%%%%%%%%%%%%%%%%%%%%%%%%%%%%%%%%%%%%%%%
%%%%%%%%%%%%%%%%%%%%%%%%%%%%%%%%%%%%%%%%%%%%%%%%%%%%%%%%%%%%%%%%%%%%%%%%%%%%
%\vspace{-5cm}
%%%%%%%%%%%%%%%%%%%%%%%%%%%%%%%%%%%%%%%%%%%%%%%%%%%%%%%%%%%%%%%%%%%%%%%%%%%%
%%%%%%%%%%%%%%%%%%%%%%%%%%%%%%%%%%%%%%%%%%%%%%%%%%%%%%%%%%%%%%%%%%%%%%%%%%%%
%%%%%%%%%%%%%%%%%%%%%%%%%%%%%%%%%%%%%%%%%%%%%%%%%%%%%%%%%%%%%%%%%%%%%%%%%%%%

\section{Dirac-Majorana neutrino mass term within seesaw in LRSM}
\label{app:B}

In this Appendix the Dirac-Majorana neutrino mass term associated with the see-saw
mass mechanism within the LRSM and particular case of neutrino mixing
given in Eq. (\ref{specmixing}) is presented. We have 
\begin{eqnarray}
\mathcal{L}_{\mathrm{D} + \mathrm{M}} &=& -\frac{1}{2}
\begin{pmatrix}
\overline{\nu_L'^C} & \overline{\nu_R'}
\end{pmatrix}
\begin{pmatrix}
M_L & M_{\mathrm{D}} \\
M_{\mathrm{D}}^{\mathrm{T}} & M_R
\end{pmatrix}
\begin{pmatrix}
\nu_L' \\
\nu_R'^C
\end{pmatrix}
+ \mathrm{H.c.} \nonumber \\
&=& -\frac{1}{2} \sum_{i = 1}^3 (m_i \overline{\nu_i} \nu_i + M_i \overline{N_i} N_i).
\end{eqnarray}
Here, $\nu_L' = \left( \nu_{eL}', \, \nu_{\mu L}', \, \nu_{\tau L}' \right)^{\mathrm{T}}$ and $\nu_R' = \left( \nu_{eR}', \, \nu_{\mu R}', \, \nu_{\tau R}' \right)^{\mathrm{T}}$ are three-component columns of the active left-handed $\nu_{\alpha L}'$ and sterile right-handed $\nu_{\alpha R}'$ ($\alpha = e, \, \mu, \, \tau$) flavor-neutrino fields, respectively. The elements of the $6 \times 6$ Dirac-Majorana mass matrix $\mathcal{M}$ are can be
calculated as follows:
\begin{eqnarray}
\mathcal{M} &=&
\begin{pmatrix}
M_L & M_D \\
M_D^{\mathrm{T}} & M_R
\end{pmatrix} \nonumber\\
&=&
\begin{pmatrix}
U & \zeta \mathds{1} \\
-\zeta \mathds{1} & U^{\dagger}
\end{pmatrix}^*
\begin{pmatrix}
m & 0 \\
0 & M
\end{pmatrix}^*
\begin{pmatrix}
U & \zeta \mathds{1} \\
-\zeta \mathds{1} & U^{\dagger}
\end{pmatrix}^{\dagger}
\end{eqnarray}
Here, $m$ and $M$ stand for diagonal $3 \times 3$ mass matrices $m = diag(m_1, m_2, m_3)$ and $M = diag(M_1, M_2, M_3)$, respectively.
By assuming the see-saw relation $m_i \sim \zeta^2 M_i$ (i=1,2,3) among light and heavy neutrino masses  the elements of 
$\mathcal{M}$ expressed in terms 3 mixing angles $\theta_{12}$, $\theta_{13}$, $\theta_{23}$, 3 CP phases $\alpha_1$, $\alpha_2$, $\delta$,
6 neutrino masses and the seesaw parameter $\zeta$ are given by
\begin{eqnarray}
  (M_L)_{ee} &=& \zeta^2 M_1 + c_{12}^2 c_{13}^2 e^{-i2\alpha_1} m_1  \nonumber\\
  &&  + s_{12}^2 c_{13}^2 e^{-i2\alpha_2} m_2 + s_{13}^2 e^{i2\delta} m_3, 
  \nonumber\\
  (M_L)_{e \mu} &=& -c_{12} c_{13} (s_{12} c_{23} + c_{12} s_{13} s_{23} e^{-i\delta}) e^{-i2\alpha_1} m_1 \nonumber\\
  &&  + s_{12} c_{13} (c_{12} c_{23} - s_{12} s_{13} s_{23} e^{-i\delta}) e^{-i2\alpha_2} m_2 \nonumber\\
  && + s_{13} c_{13} s_{23} e^{i\delta} m_3,  
  \nonumber\\
  (M_L)_{e \tau} &=& c_{12} c_{13} (s_{12} s_{23} - c_{12} s_{13} c_{23} e^{-i\delta}) e^{-i2\alpha_1} m_1 \nonumber\\
  && - s_{12} c_{13} (c_{12} s_{23} + s_{12} s_{13} c_{23} e^{-i\delta}) e^{-i2\alpha_2} m_2 \nonumber\\
  && + s_{13} c_{13} c_{23} e^{i\delta} m_3,
  \nonumber\\
  (M_L)_{\mu \mu} &=&  (s_{12} c_{23} + c_{12} s_{13} s_{23} e^{-i\delta})^2 e^{-i2\alpha_1} m_1 \nonumber\\
  && + (c_{12} c_{23} - s_{12} s_{13} s_{23} e^{-i\delta})^2 e^{-i2\alpha_2} m_2 \nonumber \\
  && + c_{13}^2 s_{23}^2 m_3 + \zeta^2 M_2,
  \nonumber\\
  (M_L)_{\mu \tau} &=& (-s_{12} s_{23} + c_{12} s_{13} c_{23} e^{-i\delta})\times \nonumber\\
  && (s_{12} c_{23} + c_{12} s_{13} s_{23} e^{-i\delta}) e^{-i2\alpha_1} m_1 \nonumber\\
  && - (c_{12} s_{23} + s_{12} s_{13} c_{23} e^{-i\delta})\times \nonumber\\
  && (c_{12} c_{23} - s_{12} s_{13} s_{23} e^{-i\delta}) e^{-i2\alpha_2} m_2 \nonumber\\
  && + c_{13}^2 s_{23} c_{23} m_3,
  \nonumber
\end{eqnarray} 
\begin{eqnarray}    
  (M_L)_{\tau \tau} &=& (s_{12} s_{23} - c_{12} s_{13} c_{23} e^{-i\delta})^2 e^{-i2\alpha_1} m_1 \nonumber\\
  && + (c_{12} s_{23} + s_{12} s_{13} c_{23} e^{-i\delta})^2 e^{-i2\alpha_2} m_2 \nonumber\\
  && + c_{13}^2 c_{23}^2 m_3 + \zeta^2 M_3, \label{mlpart}
\end{eqnarray} 

\begin{eqnarray}
  (M_D)_{ee} &=&  \zeta [-c_{12} c_{13} e^{-i\alpha_1} m_1 + c_{12} c_{13} e^{i\alpha_1} M_1],\nonumber\\
(M_D)_{e \mu} &=& \zeta [-s_{12} c_{13} e^{-i\alpha_2} m_2 + s_{12} c_{13} e^{i\alpha_2} M_1],\nonumber\\
(M_D)_{e \tau} &=& \zeta [-s_{13} e^{i\delta} m_3 + s_{13} e^{-i\delta} M_1], \nonumber\\
  (M_D)_{\mu e} &=& \zeta [(s_{12} c_{23} + c_{12} s_{13} s_{23} e^{-i\delta}) e^{-i\alpha_1} m_1 \nonumber\\
    && - (s_{12} c_{23} + c_{12} s_{13} s_{23} e^{i\delta}) e^{i\alpha_1} M_2], \nonumber\\
  (M_D)_{\mu \mu} &=& \zeta [-(c_{12} c_{23} - s_{12} s_{13} s_{23} e^{-i\delta}) e^{-i\alpha_2} m_2 \nonumber\\
    && + (c_{12} c_{23} - s_{12} s_{13} s_{23} e^{i\delta}) e^{i\alpha_2} M_2], \nonumber\\
  (M_D)_{\mu \tau} &=& \zeta [-c_{13} s_{23} m_3 + c_{13} s_{23} M_2], \nonumber\\
  (M_D)_{\tau e} &=& \zeta [-(s_{12} s_{23} - c_{12} s_{13} c_{23} e^{-i\delta}) e^{-i\alpha_1} m_1 \nonumber\\
    && + (s_{12} s_{23} - c_{12} s_{13} c_{23} e^{i\delta}) e^{i\alpha_1} M_3], \nonumber\\
  (M_D)_{\tau \mu} &=& \zeta [(c_{12} s_{23} + s_{12} s_{13} c_{23} e^{-i\delta}) e^{-i\alpha_2} m_2 \nonumber\\
    && - (c_{12} s_{23} + s_{12} s_{13} c_{23} e^{i\delta}) e^{i\alpha_2} M_3], \nonumber\\
  (M_D)_{\tau \tau} &=& \zeta [-c_{13} c_{23} m_3 + c_{13} c_{23} M_3], 
  \label{mdpart}    
\end{eqnarray}

\begin{eqnarray}
  (M_R)_{ee} &=& \zeta^2 m_1 + c_{12}^2 c_{13}^2 e^{i2\alpha_1} M_1 \nonumber\\
  && + (s_{12} c_{23} + c_{12} s_{13} s_{23} e^{i\delta})^2 e^{i2\alpha_1} M_2 \nonumber\\
  && + (s_{12} s_{23} - c_{12} s_{13} c_{23} e^{i\delta})^2 e^{i2\alpha_1} M_3, \nonumber\\
  (M_R)_{e \mu} &=& s_{12} c_{12} c_{13}^2 e^{i(\alpha_1+\alpha_2)} M_1 \nonumber\\
  && - (s_{12} c_{23} + c_{12} s_{13} s_{23} e^{i\delta})\times \nonumber\\
  &&   (c_{12} c_{23} - s_{12} s_{13} s_{23} e^{i\delta}) e^{i(\alpha_1+\alpha_2)} M_2 \nonumber\\
  && - (s_{12} s_{23} - c_{12} s_{13} c_{23} e^{i\delta})\times \nonumber\\
  &&   (c_{12} s_{23} + s_{12} s_{13} c_{23} e^{i\delta}) e^{i(\alpha_1+\alpha_2)} M_3, \nonumber\\
  (M_R)_{e \tau} &=& c_{12} s_{13} c_{13} e^{-i\delta} e^{i\alpha_1} M_1 \nonumber\\
&&  - c_{13} (s_{12} s_{23} c_{23} + c_{12} s_{13} s_{23}^2 e^{i\delta}) e^{i\alpha_1} M_2 \nonumber\\
&&  + c_{13} (s_{12} s_{23} c_{23} - c_{12} s_{13} c_{23}^2 e^{i\delta}) e^{i\alpha_1} M_3, \nonumber\\
  (M_R)_{\mu \mu} &=& \zeta^2 m_2 + s_{12}^2 c_{13}^2 e^{i2\alpha_2} M_1 \nonumber\\
  && + (c_{12} c_{23} - s_{12} s_{13} s_{23} e^{i\delta})^2 e^{i2\alpha_2} M_2 \nonumber\\
  && + (c_{12} s_{23} + s_{12} s_{13} c_{23} e^{i\delta})^2 e^{i2\alpha_2} M_3,\nonumber\\
  (M_R)_{\mu \tau} &=& s_{12} s_{13} c_{13} e^{-i\delta} e^{i\alpha_2} M_1 \nonumber\\
  && + c_{13} (c_{12} s_{23} c_{23} - s_{12} s_{13} s_{23}^2 e^{i\delta}) e^{i\alpha_2} M_2 \nonumber\\
  &&- c_{13} (c_{12} s_{23} c_{23} + s_{12} s_{13} c_{23}^2 e^{i\delta}) e^{i\alpha_2} M_3, \nonumber\\ 
  (M_R)_{\tau \tau} &=&   s_{13}^2 e^{-i2\delta} M_1 + c_{13}^2 s_{23}^2 M_2\nonumber\\
  &&  + c_{13}^2 c_{23}^2 M_3 + \zeta^2 m_3.
  \label{mrpart}
\end{eqnarray}   
We note that due to the seesaw relation $m_i \sim \zeta^2 M_i$  terms $\zeta m_i$ and $\zeta^2 m_i$
entering elements of matrices $M_D$ and $M_R$, respectively, can be safely neglected unlike
terms $\zeta^2 M_i$ appearing in the diagonal elements of the $M_L$ matrix.

\end{document}